\documentclass[10pt,journal]{IEEEtranTCOM}

\normalsize

\usepackage{amssymb}
\usepackage{cite}

\ifCLASSINFOpdf
\usepackage[pdftex]{graphicx}
\DeclareGraphicsExtensions{.pdf,.jpeg,.png}
\else
\fi

\usepackage[cmex10]{amsmath}
\usepackage{algorithm,algorithmic}

\usepackage{array}
\newcolumntype{C}[1]{>{\centering\let\newline\\\arraybackslash\hspace{0pt}}m{#1}}

\usepackage[tight,footnotesize]{subfigure}

\usepackage{stfloats}
\usepackage{multirow}
\usepackage{booktabs}
\usepackage[flushleft]{threeparttable}
\usepackage{amssymb}
\usepackage{setspace}
\usepackage{stackrel}
\usepackage{wasysym}
\usepackage{enumerate}
\usepackage[export]{adjustbox}

\newtheorem{theorem}{Theorem}

\DeclareMathSizes{10}{9}{7}{6}
\IEEEaftertitletext{\vspace{-2\baselineskip}}

\begin{document}
%
\title{Raptor Codes for Higher-Order Modulation Using a Multi-Edge Framework}
\author{Sachini~Jayasooriya,
	Mahyar~Shirvanimoghaddam,  Lawrence~Ong,
	and~Sarah~J.~Johnson}

\maketitle

\begin{abstract}
In this paper, we represent  Raptor codes as  multi-edge type low-density parity-check (MET-LDPC) codes, which  gives a general framework to  design them for higher-order modulation using  MET density evolution. We then propose an efficient Raptor code design method for higher-order modulation,  where we  design distinct degree distributions for distinct bit levels. We consider a joint decoding  scheme based on  belief propagation for Raptor codes and  also derive an exact expression for the stability condition. In several examples, we demonstrate that the higher-order modulated Raptor codes designed  using the multi-edge framework outperform previously reported higher-order modulation codes in literature. 
\end{abstract}



%
\IEEEpeerreviewmaketitle

\vspace*{-1em}
\section{Introduction}
The design of  Raptor codes with binary modulation  has been well investigated in literature~\cite{etesami-TransIT2006-raptor,cheng-TransComm2009-design}. In contrast, there has been little progress on  universal design methods for Raptor codes with bandwidth efficient higher-order modulation, and a complete analysis is still missing.  
Existing work applying Raptor codes to higher-order modulation~\cite{barron-MILCOM2009-global,venkiah-phd2012-analysis} have  considered the  bit-interleaved coded modulation scheme (BICM)~\cite{caire-1998-bit,hou-IT2003-capacity}, which uses a single binary code to protect all bits in the  signal constellation. 

However, it has been shown~\cite{zhang-THESIS-2011} that the BICM scheme, which averages over the different bit-channel reliabilities, may not show the optimal  performance if the  degree distribution is irregular, which is the case for  Raptor codes. 
In contrast, the multilevel coded (MLC) modulation~\cite{wachsmann-IT1999-multilevel,hou-IT2003-capacity} 
protects different channel bit levels using different binary codes.   Zhang~\emph{et al.}~\cite{zhang-THESIS-2011}  designed multi-edge type low-density parity-check (MET-LDPC) codes for MLC modulation for 16-quadrature amplitude modulation (16-QAM) and 
showed that the spectral efficiency can be improved when the difference in the bit-channel reliabilities are incorporated into the  code design. 
Further, Hou~\emph{et al.}~\cite{hou-IT2003-capacity}  showed that the non-binary higher-order modulation channel can be decomposed into  equivalent binary-input component channels for each individual bit level in the  signal constellation, which simplifies the analysis and design of the MLC and BICM modulation schemes. Density evolution~\cite{richardson-BOOK2008-modern}   can then be used to design  codes for each   equivalent binary-input component channel.

In this work, we propose a  general design framework for Raptor codes for higher-order modulation using a multi-edge framework. Since the multi-edge framework allows for differences in the bit channel reliabilities in the code ensemble, this enables us to design  Raptor codes using the MLC scheme and also  to perform a comprehensive analysis of the performance of the Raptor code with higher-order modulation using MET density evolution (MET-DE).  The innovation of this work is that  we simultaneously optimize distinct Raptor code degree distributions for distinct equivalent binary-input component channels using the multi-edge framework. This helps to improve the performance in higher-order modulation systems as we are incorporating the difference in the bit-channel reliabilities into the code design.

\vspace*{-1em}
\section{Target system model}

In this work, we focus on  Gray-labelled 16-QAM, which is equivalent to Gray-labelled 4-pulse-amplitude
modulation (4-PAM) mapped independently to  in-phase and quadrature components, and consider a higher-order MLC modulation scheme  and   parallel independent decoding  strategy~\cite{hou-IT2003-capacity}.

Consider a binary codeword, $c$ with code length $n$.  We can partition  $c$ into $q$-tuples.  Let the bit position in each $q$-tuple, $a$, where $a \in \mathcal{A}$ and $\mathcal{A} \in \{0,1\}^q$, be indexed by $i, i=1,\dots,q$, where $i=1$ represents the  left-most bit position.  $\mathcal{A}^i_b$ represents the subset of all elements of $\mathcal{A}$ having value $b \in \{0,1\}$ in bit position $i$.  Then a $Q$-ary modulation scheme, where $Q = 2^q$, maps each $a \in \mathcal{A}$, to a  modulated symbol, $x_a$, from a signal constellation, $\mathcal{X} \subset \mathbb{R}$, according to the Gray labelling scheme, and transmits over an additive white Gaussian noise (AWGN) channel.   Let $y$ denote the received symbol, which is the noisy version of $x_a$.  At the receiver side, an  a posteriori probability (APP) module uses $y$ to compute the log-likelihood ratio (LLR) for the coded bits $a_i, i=1,\dots, q$ as follows:
\begin{align*}
v_i &= \log \frac{\Pr (a_i = 0| Y=y)}{\Pr (a_i = 1| Y=y)}  
= \log \frac{\sum_{a \in \mathcal{A}^i_0} p ( Y=y |X=x_a)}{\sum_{a \in \mathcal{A}^i_1} p (Y=y|X=x_a)} .
\end{align*} 
The conditional density function, $p(Y=y|X=x_a)$, for  an AWGN channel  output $y$ is given by
\begin{align*}
p(Y=y|X=x_a) &= \frac{1}{\sqrt{2 \pi \sigma^2}} \exp \left( -\frac{(y-x_a)^2}{2 \sigma^2}\right),	
\end{align*}
where $\sigma^2$ is the variance of the AWGN.

One of the difficulties in the design of bandwidth efficient systems with higher-order modulations is that  equivalent binary input channels are not necessarily symmetric. Therefore the performance analysis and the code design  using density evolution  is complicated in this case, as it is not valid to assume that the all-zero codeword is transmitted  at each bit level. To address this problem,  Hou~\emph{et al.}~\cite{hou-IT2003-capacity} introduced the concept of an independent and identically distributed (iid) channel adapter, which produces an equivalent symmetric binary input channel with the same capacity.  We  incorporate this technique into the target system and design Raptor codes for  higher-order modulation using the standard MET-DE method~\cite{richardson-BOOK2008-modern}. 

Note that the extension of  the proposed method to other higher-order modulations is straightforward. 

\section{Representation of Raptor codes in the multi-edge framework for 16-QAM }

Consider a Raptor code, which is a  serial concatenation of an inner Luby transform (LT) code with degree distribution $\Omega(x) = \sum_d \Omega_d x^d$ and a precode, usually a high-rate LDPC code. For a Raptor code, first, the $k$ bit information vector is  encoded by a precode of   rate $\mathcal{R}_\mathrm{pre} = k/n$  to generate $n$  coded symbols, referred to as input bits.  Then  an LT encoder, linearly transforms this input bits into an infinite stream of LT coded symbols, referred to as output bits, which is transmitted through a channel.  
The realized rate of the LT code is defined as $\mathcal{R}_\mathrm{LT} = n/\bar{m}$, where $\bar{m}$ is the average number of LT coded symbols required for a successful decoding.

\begin{figure}[!t]
	\centering
	\includegraphics[width=0.8\linewidth]{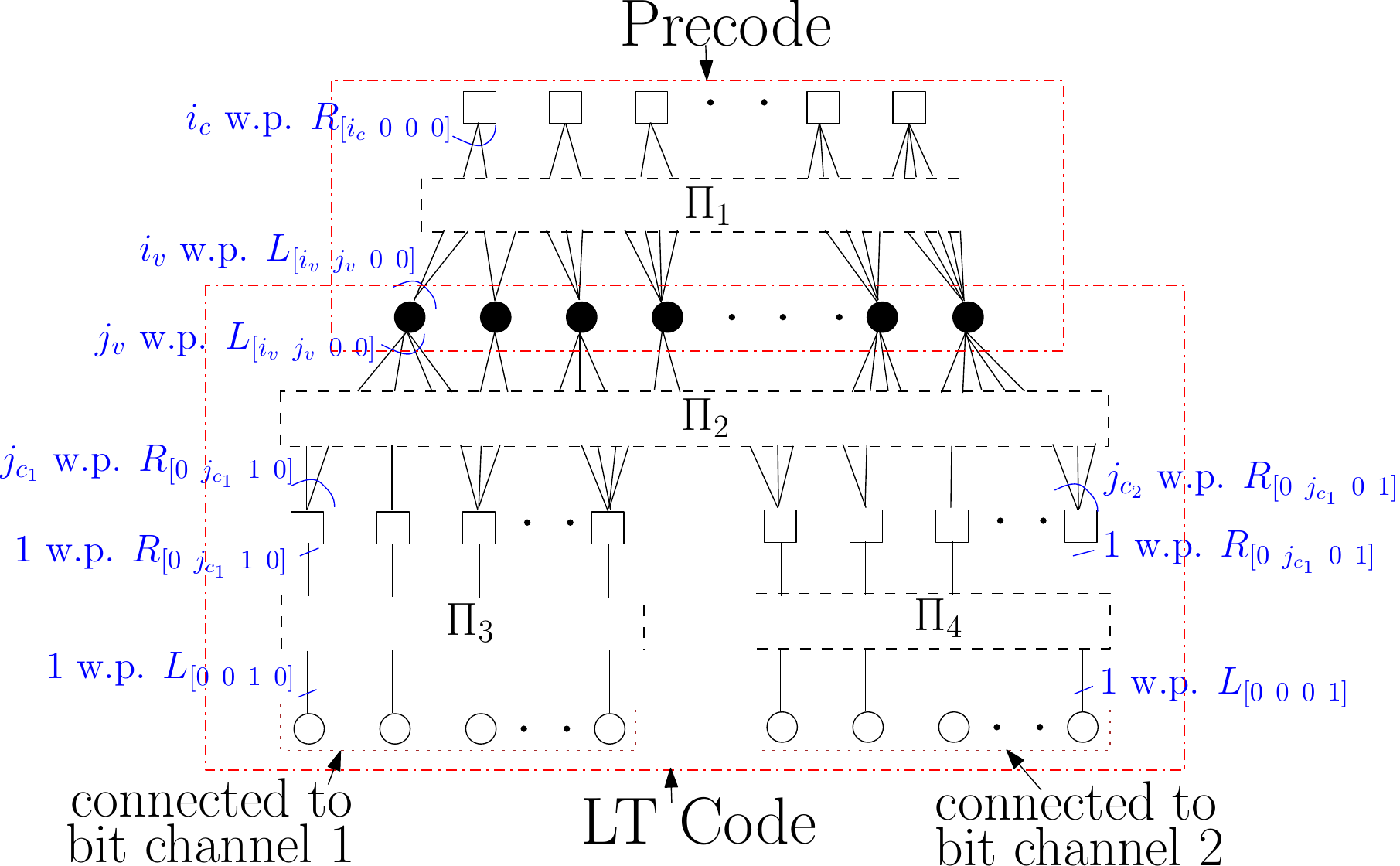}
	\caption{Graphical representation of a 16-QAM modulated Raptor code ensemble as a MET-LDPC code ensemble, where    
		\textquoteleft$\Circle$\textquoteright, \textquoteleft$\CIRCLE$\textquoteright, and  \textquoteleft$\Square$\textquoteright~ respectively represent unpunctured variable nodes, punctured variable nodes and  check nodes. \textquoteleft$\Pi_i$\textquoteright~ and $p_j$ respectively represent the edge-type $i$ and   the probability of having degree $j$ nodes.} 
	\label{fig:Raptor_code_in_MET_detail_higher}
\end{figure}

Now we  describe the new representation of 16-QAM modulated Raptor codes as MET-LDPC codes using a multi-edge framework.   Since we consider the MLC scheme, we assign one edge-type to each distinct bit level and design distinct LT degree distributions, $\Omega^{(i)}(x), i=1,2,\dots, q$, for distinct bit levels using the multi-edge framework.   
Since a Gray-labelled 16-QAM is equivalent to two Gray-labelled 4-PAM, there are only two distinct bit-channel densities.  The Tanner graph of a 16-QAM modulated Raptor code ensemble in the multi-edge framework  is shown in Fig.~\ref{fig:Raptor_code_in_MET_detail_higher}. This Tanner graph  has 4 edge-types, where  $\Pi_1$ represents the precode and $\Pi_2$ represents the LT code. $\Pi_3$ and $\Pi_{4}$ are used to input two distinct bit-channel densities of 16-QAM into the LT code. 

The 16-QAM modulated Raptor code ensemble using the multi-edge framework can be represented by  two node-perspective multinomials associated with variable nodes and check nodes  as follows:  
\begin{multline}
L(\boldsymbol{r},\boldsymbol{x}) =	r_0~\sum_{i_v=1}^{i_{v_{\max}}}~\sum_{j_v=1}^{{j_{v_{\max}}}}L_{[i_v~j_v~0~0]}~x_1^{i_v}~x_2^{j_v} ~+~ \\r_1~L_{[0~0~1~0]}~x_3 ~+~  r_{2}~L_{[0~0~0~1]}~x_{4} ,
\label{eq : raptor_varaible_higer} 	
\end{multline}
\begin{multline}
R(\boldsymbol{x}) = \sum_{i_c=2}^{{i_{c_{\max}}}} R_{[i_c~0~0~0]}~x_1^{i_c} ~+~ \sum_{j_{c_1}=1}^{{j_{c_{1_{\max}}}}}R_{[0~j_{c_1}~1~0]}~x_2^{j_{c_1}}~x_3 ~+~ \\ \sum_{j_{c_2}=1}^{{j_{c_{2_{\max}}}}}R_{[0~j_{c_2}~0~1]}~x_2^{j_{c_2}}~x_4,
\label{eq : raptor_check_higer}	
\end{multline}
where $\boldsymbol{x} = [x_1,x_2,x_3,x_4]$ corresponds to each edge-type in the Tanner graph and  $x_i^{k}$ is used to indicate the number of edges of the $i$th edge-type connected to  a particular  node. The vector $\boldsymbol{r} = [r_0,r_1,r_2]$ associated with each variable node is used to denote  bit-channel reliabilities.  In the multi-edge framework,   input bits of the Raptor code are considered as punctured variable nodes (i.e., which are not transmitting over the channel) and  denoted  by  $r_0$.  The output bits of the Raptor code are considered as unpunctured variable nodes (i.e., which are transmitted over  two different channels) and denoted  by  $r_1, r_2$.  Finally,   $L_{[i_v~j_v~k_v~m_v]}$ and $R_{[i_c~j_c~k_c~m_c]}$ are used to represent   the fraction of variable nodes of type ${[i_v~j_v~k_v~m_v]}$ and the fraction of check nodes of type ${[i_c~j_c~k_c~m_c]}$, where  fractions are relative to the number of transmitted variable nodes and $[i~j~k~m]$  denote the  number of edges, from edge-types 1 to 4, connected to that node.

We add the following  constraints into  (\ref{eq : raptor_varaible_higer}) and (\ref{eq : raptor_check_higer}) to impose the Raptor code structure into the multi-edge parametrization:
\begin{align}
&\sum_{i_v=1}^{i_{v_{\max}}}~\sum_{j_v=1}^{j_{v_{\max}}}L_{[i_v~j_v~0~0]}  = \mathcal{R}_{\mathrm{LT}}, \label{c1} \\
&\sum_{i_c=2}^{{i_{c_{\max}}}} R_{[i_c~0~0~0]} = \mathcal{R}_{\mathrm{LT}}(1-\mathcal{R}_{\mathrm{pre}}) ,\label{c2}\\
&L_{[0~0~1~0]} =  L_{[0~0~0~1]} = 0.5, \label{c3}\\
&\sum_{j_{c_1}=1}^{{j_{c_{\max}}}} R_{[0~j_{c_1}~1~0]} = \sum_{j_{c_2}=1}^{{j_{c_{\max}}}} R_{[0~j_{c_2}~0~1]} = 0.5. \label{c4}
\end{align}
Constraints (\ref{c1}) and (\ref{c2})  are used to impose the rates of LT  and precode components into the MET-LDPC code~\cite{jayasooriya2016Raptor}. Constraints  (\ref{c3}) and (\ref{c4}) are used to satisfy  constraints on the total number of transmitted bits as fractions of code length.

In this work, we design distinct LT degree distributions for distinct bit levels for the 16-QAM modulated Raptor codes using the multi-edge framework.   
We can compute these LT degree distributions from  (\ref{eq : raptor_check_higer}) as follows: 
\begin{align}
\Omega^{(1)}(x) &=  \sum_{j_{c_1}=1}^{{j_{c_{1_{\max}}}}}R_{[0~j_{c_1}~1~0]}~x^{j_{c_1}}, \label{omega1} \\
\Omega^{(2)}(x) &=  \sum_{j_{c_2}=1}^{{j_{c_{2_{\max}}}}}R_{[0~j_{c_2}~0~1]}~x^{j_{c_2}}, \label{omega2} 
\end{align}
where $\Omega^{(i)}(x)$  is the LT degree distribution related to bit level $i$.
Note that these polynomials, $\Omega^{(1)}(x)$ and $\Omega^{(2)}(x)$,  are not valid probability mass functions as their coefficients do not sum to one.

The  rate efficiency of a $Q$-ary modulated Raptor code ensemble at signal-to-noise ratio (SNR), $\gamma$, can be computed as 
\begin{align}
\eta(\gamma) &= \left(\frac{ \mathcal{R}(\gamma) }{C(\gamma)}\right) \log_2 Q ,  \label{Rptor_rate_eff_higer}
\end{align}
where $C(\gamma)$ is the capacity of the AWGN channel with $Q$-ary modulation at SNR, $\gamma$~\cite{hou-IT2003-capacity} and  $\mathcal{R}(\gamma)$ is the designed rate of the Raptor code, which can be computed using~{\protect\cite[page 383]{richardson-BOOK2008-modern}}
\begin{align}
\mathcal{R}(\gamma) &= L(\boldsymbol{1},\boldsymbol{1}) - R(\boldsymbol{1}),
\label{code rate}
\end{align} 
where $\boldsymbol{1}$ denotes a vector of all $1$'s with the length determined by the context.  

\section{Design of  16-QAM modulated  Raptor codes  using the multi-edge framework}
\subsection{Decoding of  Raptor codes in the multi-edge framework}
Generally, the decoding process of Raptor codes consists of a series of decoding attempts, which runs the LT decoder for a predetermined number of belief propagation (BP) decoding iterations, which is followed separately by the precode decoder.  In this work, we instead consider a joint decoding scheme based on the BP decoding for the decoding of Raptor codes, where both component codes (LT code and precode) are decoded in parallel and extrinsic information (i.e., the information from previous BP decoding iteration) is exchanged between the decoders~\cite{jayasooriya2016Raptor}.   
We consider the stability of  $Q$-ary  modulated Raptor codes and derived an exact expression for the stability condition when  Raptor codes are decoded with  joint decoding  using the multi-edge framework, which is given in  Theorem~\ref{Stability_Raptor}.

\begin{theorem}\label{Stability_Raptor}
Consider a  Q-ary modulated Raptor code  with  joint decoding  based on  the BP decoding  using the multi-edge framework. On the AWGN channel, the stability condition is given by	
	\begin{align*}			 
	\sum_{j\geq 1} \lambda_{[2~j]}(y_2)^j~\rho_1'(1) &\leq 1 ,
	\end{align*}
where $\lambda_{[2~j]}$ gives the fraction of degree-two variable nodes in the precode  component	and $\rho_1'(1)$ is the derivative of  $\rho_1(x)$, which is the edge-perspective check node degree distribution for the edge-type 1 ($\Pi_1$), when $x=1$.   $y_2 = \frac{1}{q} \sum_{i=1}^{q}x_0^{(i)} $, where
$x_0^{(i)}$  is the Bhattacharya constant~{\protect\cite[page 202]{richardson-BOOK2008-modern}} associated with the channel density corresponding to bit level $i$ and $q = \log_2 Q$.	 
\end{theorem}
\begin{IEEEproof}
	See Appendix~\ref{appendix_A}	
\end{IEEEproof}
Note that the stability  condition of a $Q$-ary modulated Raptor codes is similar to the stability condition of   binary modulated Raptor codes given in~\cite{jayasooriya2016Raptor}  having multiple channel inputs.

\subsection{Raptor code optimization for 16-QAM} \label{sec-raptor design}
The aim of the Raptor code optimization in the multi-edge framework is to find the largest possible realized rate, $\mathcal{R}(\gamma)$, for a given SNR, $\gamma$, over the check node degree distribution, $R(\boldsymbol{x})$, under the constraints given in (\ref{c1}) to (\ref{c4}). Then we can find the corresponding variable node degree distribution,   $L(\boldsymbol{r},\boldsymbol{x})$, using the rate constraint given in  (\ref{code rate}) and the socket count equality constraint which is given by

$L_{x_i}(\boldsymbol{1},\boldsymbol{1}) = \frac{d L(\boldsymbol{r},\boldsymbol{x})}{d{x_i}} \biggr|_{\boldsymbol{r}=\boldsymbol{1},\boldsymbol{x}=\boldsymbol{1}} $
and 
$R_{x_i}(\boldsymbol{1}) = \frac{d R(\boldsymbol{x})}{d{x_i}} \biggr|_{\boldsymbol{x}=\boldsymbol{1}}$. 

The optimization problem for 16-QAM modulated Raptor codes for a given SNR, $\gamma$, in the multi-edge framework  can be summarized as follows:

 \hspace*{0.2cm} ${\mathrm{Maximize}} 
 \hspace*{0.3cm} \mathcal{R}(\gamma) $
\begin{align*}
\begin{aligned}
& \mathrm{s.t.}
& & (i)~ \sum_{i_v=1}^{i_{v_{\max}}}~\sum_{j_v=1}^{j_{v_{\max}}}L_{[i_v~j_v~0~0]}  = \mathcal{R}_{\mathrm{LT}}, \\
&&& (ii)~ \sum_{i_c=2}^{{i_{c_{\max}}}} R_{[i_c~0~0~0]} = \mathcal{R}_{\mathrm{LT}}(1-\mathcal{R}_{\mathrm{pre}}) , \\
&&& (iii)~ L_{[0~0~1~0]} =  L_{[0~0~0~1]} = 0.5,\\ 
&&& (iv)~ \sum_{j_{c_1}=1}^{{j_{c_{\max}}}} R_{[0~j_{c_1}~1~0]} = \sum_{j_{c_2}=1}^{{j_{c_{\max}}}} R_{[0~j_{c_2}~0~1]} = 0.5 \\ 
&&& (v)~ L_{x_i}(\boldsymbol{1},\boldsymbol{1}) = R_{x_i}(\boldsymbol{1}),\\
&&& (vi)~ \max_j\{\epsilon_j\} < \epsilon^*. 
\end{aligned}
\end{align*}
where   $\epsilon_j$ and $\epsilon^*$ respectively denote  the bit error rate (BER)  on the $j$th variable node and target maximum BER.  
We use combined optimization method proposed in~\cite{jayasooriya2016Joint} (adaptive range method~\cite{jayasooriya2016Joint} to optimize the node fractions and differential evolution~\cite{storn-book1997-differential}  to optimize the node degrees)  to optimize    Raptor code ensembles, which are represented in the multi-edge framework, and compute the realized rate $\mathcal{R}(\gamma)$ using MET-DE.

\vspace*{-1.5em}
\section{Results and discussion}
\begin{figure}[!t]
	\centering
	\includegraphics[width=0.9\linewidth]{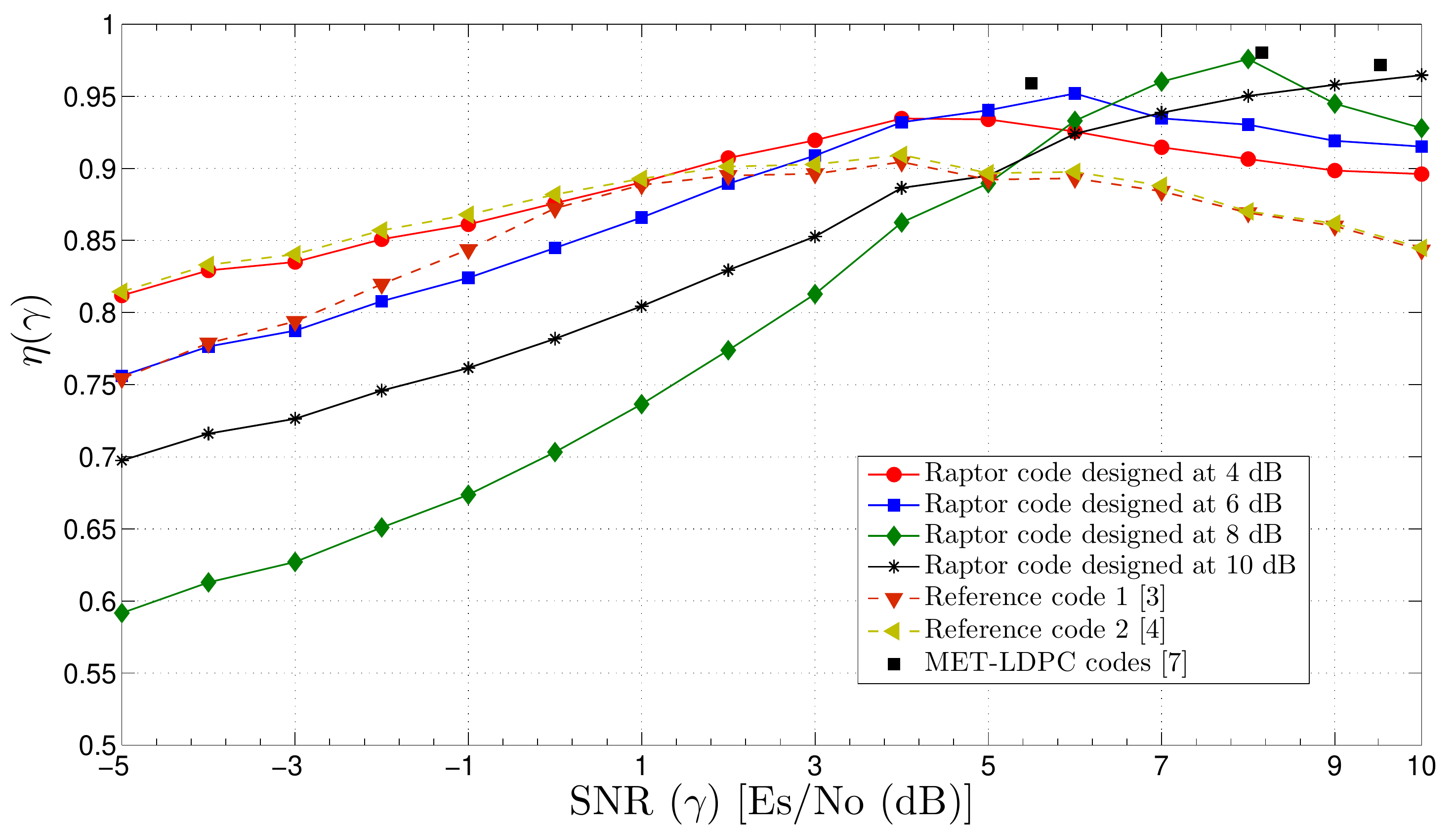}
	\vspace*{-1em}
	\caption{Rate efficiency  results computed using MET-DE for the optimized Raptor codes shown in Table.~\ref{tab:opt_codes}.}
	\label{fig:figrateeff}
\end{figure}

We formulate the optimization problem as per Section~\ref{sec-raptor design}  and optimize  Raptor codes for 16-QAM using MLC modulation. For the Raptor code design we  set
the precode to a (3,60)-regular LDPC code with a rate of 0.95, and  set the maximum output node degree to be 50. The optimized LT degree distributions and their rate efficiencies are shown in Table~\ref{tab:opt_codes}.
In Fig.~\ref{fig:figrateeff}, we evaluate the rate efficiency performance for the optimized Raptor code ensembles for SNRs above and below the designed SNR using MET-DE. The rate efficiency performance for two reference Raptor codes from the literature~\cite{barron-MILCOM2009-global,venkiah-phd2012-analysis} are also shown in the same figure for comparison. The reference code 1~\cite{barron-MILCOM2009-global} was designed using BICM  for Gray-labelled 16-QAM  with a precode of (3,30)-regular LDPC code with a rate of 0.9.  Venkiah~\cite{venkiah-phd2012-analysis} designed Raptor codes for a  binary input AWGN channel with a precode of (3,60)-regular LDPC code and use those codes to evaluate the performance with 16-QAM. Thus we chose the Raptor degree distribution from~\cite[Tabel 2.4]{venkiah-phd2012-analysis} and use that as  reference code 2. We have also compared several MET-LDPC code ensembles designed using MLC modulation for 16-QAM in~\cite[Tabel 4.2]{zhang-THESIS-2011} in Fig.~\ref{fig:figrateeff}.  

It is clear from Fig.~\ref{fig:figrateeff} and Table~\ref{tab:opt_codes}  that Raptor codes with  appropriately designed degree distributions for distinct bit channel densities using MLC modulation can achieve better rate efficiency performance compared to the previously reported Raptor codes in the literature.
Moreover, in contrast to 16-QAM modulated MET-LDPC codes~\cite[Tabel 4.2]{zhang-THESIS-2011}, which are fixed-rate codes,   Raptor codes can achieve good rate  efficiency performance in the entire SNR range.  This shows the benefit of considering Raptor codes for the  higher-order modulation compared to traditional fixed-rate coding schemes.

\vspace*{-1em}
\section{Conclusion}
In this paper, we represented  Raptor codes as  multi-edge type low-density parity-check (MET-LDPC) codes, to  design  higher-order modulated Raptor codes using multi level coded (MLC) modulation. The MET representation enables us to design  distinct degree distributions for distinct bit channel densities for MLC modulation using MET density evolution.  In several examples, we demonstrated that the higher-order modulated Raptor codes designed  using the multi-edge framework outperform previously reported Raptor codes in literature.

\begin{table*}[!t]
	\renewcommand{\arraystretch}{1.1}
	\centering
	\caption{Optimized Raptor code degree distributions  for 16-QAM  using different designed SNRs ($\gamma_d$)} \vspace*{-1em}
	\scriptsize
	\begin{tabular}{|c|l|c|}
		\hline
		$\gamma_d$ & \multicolumn{1}{c|}{Degree distribution} & $\eta(\gamma_d)$ \\
		\hline
		\multirow{2}[4]{*}{{4 dB}} & $\Omega^{(1)}(x) = 0.0101\, x+ 0.18565\, x^2+ 0.0948\, x^3+ 0.14205\, x^4+ 0.00875\, x^5+ 0.00785\, x^9+ 0.0184\, x^{13}+ 0.03005\, x^{18}+ 0.0024\, x^{50}$ & \multirow{2}[4]{*}{\textbf{0.9345}} \\
		& $\Omega^{(2)}(x) =0.01465\, x+ 0.32605\, x^2+ 0.0493\, x^3+ 0.06465\, x^5+ 0.01435\, x^9+ 0.0017\, x^{11}+ 0.01855\, x^{13}+ 0.00905\, x^{22}+0.00175\, x^{48}$ &  \\
		\hline
		\multirow{2}[4]{*}{{6 dB}} & $\Omega^{(1)}(x) = 0.01\, x+ 0.1959\, x^2+ 0.11015\, x^3+ 0.12755\, x^4+ 0.02195\, x^5+ 0.01075\, x^9+0.0071\, x^{13}+ 0.0153\, x^{18}+ 0.00125\, x^{50}$ & \multirow{2}[4]{*}{\textbf{0.9519}} \\
		& $\Omega^{(2)}(x) =0.01375\, x+ 0.3361\, x^2+ 0.0394\, x^3+ 0.0596\, x^5+ 0.011\, x^9+ 0.0027\, x^{11}+ 0.0188\, x^{13}+ 0.01075\, x^{22}+ 0.00785\, x^{48}$ &  \\
		\hline
		\multirow{2}[4]{*}{{8 dB}} & $\Omega^{(1)}(x) =0.01295\, x+ 0.38435\, x^2+ 0.09615\, x^5+ 0.00615\, x^{13}+ 0.00005\, x^{16}+ 0.0003\, x^{22}+ 0.0001\, x^{48}$ & \multirow{2}[4]{*}{\textbf{0.9759}} \\
		& $\Omega^{(2)}(x) = 0.1641\, x^2+0.2872\, x^3+ 0.0412\, x^6+ 0.00105\, x^{11}+ 0.0061\, x^{15}+ 0.0002\, x^{19}+ 0.0001\, x^{46}$ &  \\
		\hline
		\multirow{2}[4]{*}{{10 dB}} & $\Omega^{(1)}(x) =0.21975\, x^2+ 0.05065\, x^3+ 0.2076\, x^4+ 0.0146\, x^6+ 0.00065\, x^{10}+ 0.00085\, x^{12}+ 0.0023\, x^{18}+ 0.0037\, x^{48}$ & \multirow{2}[4]{*}{\textbf{0.9646}} \\
		& $\Omega^{(2)}(x) =0.01915\, x+ 0.40395\, x^2+ 0.0493\, x^3+ 0.0156\, x^7+ 0.00045\, x^{12}+ 0.0003\, x^{13}+ 0.01125\, x^{49}$ &  \\
		\hline
	\end{tabular}%
	\vspace*{-2em}
	\label{tab:opt_codes}%
\end{table*}

\vspace*{-1em}
\appendices
\section{Proof of Theorem~\ref{Stability_Raptor}} \label{appendix_A}

As indicated in (\ref{eq : raptor_varaible_higer}), the variable node degree distribution of a Raptor code ensemble in the multi-edge framework  has degree-one variable nodes. Thus we consider the special case given in~{\protect\cite[pages 396-397]{richardson-BOOK2008-modern}} and  define nodes connected to $\Pi_1$ as the core LDPC graph for the stability analysis. The edge-perspective multinomial of the  core LDPC graph with edges connected to $\Pi_2$  can be computed using (\ref{eq : raptor_varaible_higer}) and (\ref{eq : raptor_check_higer}) as follows:
\vspace*{-1em}
\begin{align}
\lambda_1(x_1, x_2) &= \frac{L_{x_1}(\boldsymbol{r},\boldsymbol{x})}{L_{x_1}(\boldsymbol{1},\boldsymbol{1})} =  \sum_{i\geq2} \sum_{j\geq 1} \lambda_{[i~j]}~x_1^{i-1}~x_2^{j}, \label{core_VN} \\
\rho_1(x_1) &=  \frac{R_{x_1}(\boldsymbol{x})}{R_{x_1}(\boldsymbol{1})} =\sum_{i\geq 0} \rho_{i}~x_1^{i-1}, \label{core_CN}
\end{align}

The stability analysis of  Raptor codes with joint decoding using MET-DE examines the asymptotic behavior of the BP decoder when it is close to a successful decoding and gives a sufficient condition for the convergence of  the BER  to zero  as the BP decoding iteration, $\ell$, tends to infinity.   Therefore as stated in~{\protect\cite[pages 396-397]{richardson-BOOK2008-modern}} we can assume that check nodes connected to $\Pi_3$ all carrying a message with the same distribution as the bit channel 1 density  and check nodes connected to $\Pi_4$ all carrying a message with the same distribution as the bit channel 2 density, in the check-to-variable direction along the edges connected to  $\Pi_2$. Then the proof of Theorem~\ref{Stability_Raptor} is as follows: 

Let $a_i^{(\ell)}$ and $b_i^{(\ell)}$ respectively  denote the PDFs of  variable-to-check message  and check-to-variable message along $\Pi_i$  
at the $\ell$th BP decoding iteration. Let $x_i^{(\ell)}$ and  $y_i^{(\ell)}$ respectively  denote the Bhattacharyya constants associated with $a_i^{(\ell)}$ and $b_i^{(\ell)}$. 
DE equations related to variable nodes and check nodes  of the core LDPC graph can be written as follows:
\begin{align}
\label{variable_node_update_core}
a_1^{(\ell+1)} &=  \sum_{i\geq2} \sum_{j\geq 1} \lambda_{[i~j]}\left[b_1^{(\ell)}\right]^{\otimes (i-1)} 		\otimes\left[b_2^{(\ell)}\right]^{\otimes (j) }  = \lambda_1(b_1^{(\ell)},b_2^{(\ell)}),  \\
\label{check_node_update_core}
b_1^{(\ell)} &=  \sum_{i\geq 0} \rho_i\left[a_1^{(\ell)}\right]^{\boxtimes (i-1)} 	= \rho_1(a_1^{(\ell)}),
\end{align}
where  $\otimes$ denotes the variable node convolution and $\boxtimes$ denotes the check node convolution~\cite{richardson-BOOK2008-modern}.
Moreover, for a $Q$-ary modulated Raptor code, the average check-to-variable message along $\Pi_2$ at the $\ell$th decoding iteration is the average over all bit channel densities.  Thus
\begin{align*}
b_2^{(\ell)} 	&= \frac{1}{q} \sum_{i=1}^{q}a_0^{(i)} , 
\end{align*}
where $a_0^{(i)}$ is the  bit channel density corresponding to bit level $i$ and $q=\log_2Q$. Then by applying Lemma 4.63 in~{\protect\cite[pages 202]{richardson-BOOK2008-modern}} to  (\ref{variable_node_update_core}),  we get the final update rule for $x_1^{(\ell+1)}$  as follows:
\begin{align*}
x_1^{(\ell+1)} &\leq \sum_{i\geq2} \sum_{j\geq 1} \lambda_{[i~j]}~\left(1-\sum_k \rho_k (1-x_1^{(\ell)})^{k-1}\right)^{i-1}\left(y_2^{(\ell)}\right)^{j}, 
\end{align*}
where $y_2^{(\ell)} = \frac{1}{q} \sum_{i=1}^{q}x_0^{(i)}$ and 
$x_0^{(i)}$ is the bhattacharya constant  associated with  $a_0^{(i)}$. 
Furthermore, to ensure that the BER decreases throughout the BP decoding iterations condition, $x_1^{(\ell+1)}< x_1^{(\ell)}$ must hold. Thus for a successful decoding under MET-DE, we need to guarantee that  
\begin{equation}
x_1^{(\ell)} > \sum_{i\geq2} \sum_{j\geq 1} \lambda_{[i~j]}\left(1-\sum_k \rho_k (1-x_1^{(\ell)})^{k-1}\right)^{i-1}\left(y_2^{(\ell)}\right)^{j}. \label{B_4}
\end{equation}
For the decoding to be successful, inequality (\ref{B_4}) needs to be valid around  $x_1^{(\ell)}=0$. Thus taking the derivative of  (\ref{B_4}) with respect to $x_1^{(\ell)}=0$ gives us
\begin{multline*}
\lambda_1'(x_1^{(\ell)},x_0) \times \rho_1'(1-x_1^{(\ell)})\big|_{x_1^{(\ell)}=0} = \\
\sum_{j\geq 1} \lambda_{[2~j]}~(y_2)^j \times\rho_1'(1)\leq 1.
\end{multline*}	
This completes the proof.

\vspace*{-0.5em}
\footnotesize
\singlespacing

\end{document}